\documentclass[12pt,,letterpaper]{JHEP3}
\usepackage{graphicx}
\usepackage{epsfig}

\title{Note on the non-adjacent BCFW deformations}

\author{Hongbao Zhang\\
Crete Center for Theoretical Physics, Department of Physics, \\
University of Crete, 71003 Heraklion, Greece\\
\email{hzhang@physics.uoc.gr} }
 \abstract{As the first part, we provide a proof of the consistency condition by Kleiss-Kuijf relation and the bonus relation by BCJ
 relation for the non-adjacent BCFW deformations. On the
other hand, rather than appealing to field theory argument, we
provide an alternative proof of bonus relation for the non-adjacent
BCFW deformations by a purely S matrix analysis in the context of
$\mathcal{N}=4$ SYM theory.}

\begin{document}
\section{Introduction}
S matrix program is to try to construct the scattering amplitude
based on some general principles such as Poincare invariance,
analyticity, and unitarity. A great advance along this line is the
recent discovery of BCFW recursion relation, which allows us to
construct the tree level on-shell scattering amplitude from the
lower ones\cite{BCF,BCFW}. In particular, by the consistency
condition between the adjacent BCFW deformations, one can construct
the tree level scattering amplitude for massless spin-1 particles in
four dimensional Minkowski spacetime just under the good BCFW
deformation on any adjacent pair without resorting to the underlying
quantum theory of gauge fields\cite{BC,HZ,ST}.

However, there appears to be some additional information hidden in
the non-adjacent BCFW deformations. For example, in the most recent
derivation of Kleiss-Kuijf relation from BCFW recursion relation,
the consistency condition imposed by the non-adjacent BCFW
deformations has been employed, and it is also shown that BCJ
relation can be obtained by the bonus relation from the non-adjacent
BCFW deformations\cite{FHJ}.

The purpose of this note is two-fold. Firstly, we provide a proof of
the consistency condition by Kleiss-Kuijf relation and the bonus
relation by BCJ relation for the non-adjacent BCFW deformations
reversely, which thus implies that such a consistency condition and
bonus relation are essentially equivalent to Kleiss-Kuijf relation
and BCJ relation respectively. On the other hand, rather than
appealing to field theory argument\cite{AK}, we reproduce the bonus
relation for the non-adjacent BCFW deformation by a purely S matrix
analysis, where $\mathcal{N}=4$ SYM theory will be employed for
convenience.
\section{Proof of consistency condition by Kleiss-Kuijf relation}
We firstly introduce the Kleiss-Kuijf relation\cite{KK}, i.e.,
\begin{equation}
A_n(1,\{\alpha\}, i, \{\beta\})=(-1)^{n_\beta}\sum_\sigma
A_n(1,\sigma(\{\alpha\}, \{\beta^T\}),i),
\end{equation}
where the set $\{\beta^T\}$ denotes the reversed ordering of
$\{\beta\}$, $n_\beta$ is the number of elements in $\{\beta\}$, and
$\sigma$ takes over all the permutations which preserve the relative
ordering in $\{\alpha\}$ and $\{\beta^T\}$. It is noteworthy that
the Kleiss-Kuijf can actually include the color-order reversed
relation and $U(1)$ decoupling equation as its special cases.
Speaking specifically, when we set $\{\alpha\}=\varnothing$, the
Kleiss-Kuijf relation gives us the color-order reversed relation,
i.e.,
\begin{equation}
A_n(1,2,\cdot\cdot\cdot,n)=(-1)^nA_n(n,n-1,\cdot\cdot\cdot,1),
\end{equation}
on the other hand, when we set $n_\beta=1$, the Kleiss-Kuijf
relation follows the $U(1)$ decoupling equation, i.e.,
\begin{equation}
\sum_{\sigma\in cyclic}A_n(1,\sigma(2,3,\cdot\cdot\cdot,n))=0.
\end{equation}

Now let us do the BCFW deformation on the scattering amplitude
$A(1,\{\alpha\},i,\{\beta\})$ by shifting the non-adjacent pair
denoted by $[1,i]$, i.e.,
\begin{equation}
A_{[1,i]}(1,\{\alpha\},i,\{\beta\})=\sum A_L(\hat{P},
\{\beta_2\},\hat{1},\{\alpha_1\})\frac{1}{P^2}A_R(\hat{i},\{\beta_1\},-\hat{P},\{\alpha_2\}),
\end{equation}
where $\sum$ represents the summation over internal helicities, and
all possible divisions of $\{\alpha\}=\{\alpha_1\}\cup\{\alpha_2\}$
and $\{\beta\}=\{\beta_1\}\cup\{\beta_2\}$. Applying the
Kleiss-Kuijf relation to both of the left and right hand side
amplitudes, we obtain
\begin{eqnarray}
A_{[1,i]}(1,\{\alpha\},i,\{\beta\})&=&\sum(-1)^{\alpha_1+\alpha_2}A_L(\hat{P},\sigma^L(\{\beta_2\},\{\alpha_1^T\}),\hat{1})\frac{1}{P^2}
A_R(\hat{i},\sigma^R(\{\beta_1\},\{\alpha^T_2\}),-\hat{P})\nonumber\\
&=&\sum(-1)^{\alpha}A_L(\hat{P},\sigma^L(\{\beta_2\},\{\alpha_1^T\}),\hat{1})\frac{1}{P^2}
A_R(\hat{i},\sigma^R(\{\beta_1\},\{\alpha^T_2\}),-\hat{P}).\nonumber\\
\end{eqnarray}
Obviously, now the BCFW deformation on the pair becomes an adjacent
one, denoted by $(1,i)$, i.e.,
\begin{equation}
A_{[1,i]}(1,\{\alpha\},i,\{\beta\})=\sum(-1)^\alpha
A_{(1,i)}(i,\sigma(\{\beta\},\{\alpha^T\}),1)
=A(1,\{\alpha\},i,\{\beta\}),
\end{equation}
where we have used the fact that
$\sigma^L(\{\beta_2\},\{\alpha_1^T\})$, together with
$\sigma^R(\{\beta_1\}\{\alpha^T_2\})$ is the same as
$\sigma(\{\beta\},\{\alpha^T\})$ in the first step, and employed the
Kleiss-Kuijf relation in the last step. Therefore, if the
Kleiss-Kuijf relation holds for the scattering amplitude constructed
from the adjacent BCFW deformation, then the non-adjacent BCFW
deformation produces the same scattering amplitude as the adjacent
one, which completes our proof.
\section{Proof of bonus relation by BCJ relation}
Let us start with BCJ relation\cite{BCJ}, i.e.,
\begin{equation}
A_n(1,2,\{\alpha\},3,\{\beta\})=\sum_\sigma
A_n(1,2,3,\sigma(\{\alpha\},\{\beta\})) \prod_{k=4}^m {{\cal
F}(3,\sigma(\{\alpha\},\{\beta\}),1| k)\over s_{2,4,\ldots,k} }.
\label{BCJ}
\end{equation}
Here without loss of generality, we set $\{\alpha\} \equiv
\{4,5,\ldots,m-1,m\}$ and $\{\beta\} \equiv
\{m+1,m+2,\ldots,n-1,n\}$.  In addition, $\sigma$ denotes all the
permutations of the set $\{\alpha\}\cup\{\beta\}$ that maintains the
order of elements in $\{\beta\}$ . The function ${\cal F}$
associated with $k$ is given by
\begin{eqnarray}
{\cal F}(3,\sigma(\{\alpha\},\{\beta\}),1| k)
     \equiv {\cal F} (\{\rho\}| k)&=&
\left\{
\begin{array}{ll}
          \sum_{l=t_k}^{n-1} {\cal G}(k,\rho_l) & \mathbf{if} t_{k-1} < t_k\\
        - \sum_{l=1}^{t_k} {\cal G}(k,\rho_l) & \mathbf{if} t_{k-1} > t_k
\end{array} \right\}  \nonumber\\&&
 \null +
\left\{
\begin{array}{ll}
         s_{2,4,\ldots,k} & \mathbf{if} t_{k-1} < t_k< t_{k+1}\\
        -s_{2,4,\ldots,k} & \mathbf{if} t_{k-1} > t_k > t_{k+1}\\
         0 & \mathbf{else}
\end{array} \right\} .
\label{Fun}
\end{eqnarray}
Here $t_k$ is the position of $k$ in the set $\{\rho\}$, except that
we set $t_3\equiv t_5$ and $t_{m+1}\equiv 0$ as the boundary
condition once and for all. The function ${\cal G}$ is given by
\begin{equation}
{\cal G }(i,j)=\left\{
\begin{array}{ll}
        s_{i,j}  & \mathbf{if} i< j \mathbf{or} j=1,3\\
        0& \mathbf{else}
\end{array} \right\} .
\end{equation}
Finally, the kinematic invariants are defined as
\begin{eqnarray}
s_{i,j}&=&(k_i+k_j)^2,\nonumber\\
 s_{2,4,\ldots,i}&=&(k_2+k_4+\ldots+k_i)^2.
\end{eqnarray}

Apparently, to prove the bonus relation for the non-adjacent BCFW
deformations is equivalent to show the corresponding large $z$
behavior goes like $\frac{1}{z^2}$. Note that BCJ relation is
trivial in the case of $\{\alpha\}=\varnothing$. Thus in what
follows, we shall assume the set $\{\alpha\}\neq\varnothing$, which
means the BCFW deformation of the pair $[2,3]$ on the left hand side
of Eq.(\ref{BCJ}) is always a non-adjacent one. So it is sufficient
for us to show the whole product part of right hand side of
Eq.(\ref{BCJ}) is of order $\frac{1}{z}$ when $z$ goes to infinity,
as the same deformation on the adjacent pair $(2,3)$ of the
amplitude part gives the order $\frac{1}{z}$ by BCFW recursion
relation.

It is noteworthy that in each $k$th term of the product part, the
denominator always contributes order of $z$ due to the deformation
on particle $2$, while the numerator contributes order of $z$ or
order of $1$, depending on the specific condition on $t_k$. So it is
only necessary for us to show that there exists at least one
numerator which contributes order of $1$.

Firstly, let us check the case of $m=4$ where $t_3=t_5=0$ due to the
boundary condition. So by Eq.(\ref{Fun}), the contribution from the
numerator gives the desired order of $1$. Similarly, for the case of
$m=5$, if $t_3<t_4$, then the contribution from the $4$th term
follows order of $1$; on the other hand, if $t_3>t_4$, then the
contribution from the $5$th term yields order of $1$.

Now let us move on to the more general cases, i.e., $m>5$. Here we
assume that each $k$th term contributes order of $z$ if only $k\neq
m$, otherwise the proof is automatically completed by definition. By
Eq.(\ref{Fun}), such an assumption implies that
$t_{k-1}<t_k<t_{k+1}$ should hold for $k=4,5,\ldots,m-1$. In
particular, we thus have $t_{m-1}<t_m>t_{m+1}=0$. Then it follows
from Eq.(\ref{Fun}) that the contribution from the $m$th term
produces the expected order of $1$, which thus completes our proof.
\section{Proof of bonus relation by a purely S matrix analysis in the context of $\mathcal{N}=4$ SYM theory}
In what follows, to eschew the cumbersome helicity analysis, we
shall work in the context of $\mathcal{N}=4$ SYM theory, which is
well known to produce the same result for the tree amplitude in
purely gauge theory.
\subsection{Super-BCFW recursion relation in $\mathcal{N}=4$
SYM theory} In $\mathcal{N}=4$ SYM theory, we can group all on-shell
states into a super-wavefunction as\cite{DHKS,ACK}
\begin{eqnarray}
\Phi(p,\eta)&=&G^+(p)+\eta^A\Gamma_A(p)+\frac{1}{2}\eta^A\eta^BS_{AB}(p)\nonumber\\
&&+\frac{1}{3!}\eta^A\eta^B\eta^C\epsilon_{ABCD}\bar{\Gamma}^D(p)+\frac{1}{4!}\eta^A\eta^B\eta^C\eta^D\epsilon_{ABCD}G^-(p),
\end{eqnarray}
where $\eta^A$ is the Grassmann variable with $A=1,2,3,4$. Whence
the corresponding super-amplitude can be written as a function of
$(\lambda,\widetilde{\lambda},\eta)$. For example, the super-MHV
amplitude is given by\cite{Nair}
\begin{equation}
A_n=\frac{\delta^4(\sum_{i=1}^n\lambda_i\tilde{\lambda}_i)\delta^8(\sum_{i=1}^n\lambda_i\eta_i)}{\langle
1|2\rangle\langle 2|3\rangle\ldots\langle
n|1\rangle}=\frac{\delta^4(\sum_{i=1}^n\lambda_i\tilde{\lambda}_i)\delta^8(\sum_{i=1}^n\lambda_i\eta_i)}{\mathbf{cyc}(
1,2,\ldots,n)}.\label{MHV}
\end{equation}
The purely gluon amplitude can be obtained by integral over the
grassmann variable or setting it to be zero, depending the specific
helicity of gluon.

To guarantee the supersymmetric counterpart of momentum
conservation, besides the ordinary deformation on the pair $(k,l)$,
i.e.,
\begin{eqnarray}
\lambda_k(z)&=&\lambda_k+z\lambda_l,\nonumber\\
\tilde{\lambda}_l(z)&=&\tilde{\lambda}_l-z\tilde{\lambda}_k,
\end{eqnarray}
one need also do the additional deformation for $\eta$ as
\begin{equation}
\eta_l(z)=\eta_l-z\eta_k.
\end{equation}
With such a deformation, the super-BCFW recursion relation can be
expressed as\cite{ACK}
\begin{equation}
A_{k,l}=\sum_{L,R}\int d^4\eta_P
A_L[\lambda_k(z_0),-\lambda_P(z_0),\tilde{\lambda}_P(z_0),\eta_P]\frac{1}{P^2}A_R[\lambda_P(z_0),
\tilde{\lambda}_P(z_0),\eta_P,\tilde{\lambda}_l(z_0),\eta_l(z_0)].\label{BCFW}
\end{equation}
Note that the minus sign is judiciously chosen on the left hand side
$\lambda_P$ such that the left hand side momentum has the opposite
sign as the right hand side one. Such a choice also ensures the
supersymmetry.
\subsection{Super-MHV expansion in $\mathcal{N}=4$ SYM theory}
Starting from the amplitude constructed essentially by any adjacent
BCFW deformation, it has been shown without any other assumption
that MHV vertex expansion is valid for all tree amplitudes in
$\mathcal{N}=4$ SYM theory\cite{EFK1,EFK2}. Such a result has also
been generalized to super-MHV vertex expansion\cite{KN}. Speaking
specifically, the large $z$ behavior of N$^k$MHV super-amplitude
goes like $\frac{1}{z^k}$ under the all-line supershifts, i.e., \
\begin{eqnarray}
\tilde{i}(z)&=&\tilde{i}+zc_i\tilde{X},\nonumber\\
\eta_i(z)&=&\eta_i+zc_i\eta_X \label{Risager}
\end{eqnarray}
Here $c_i$ satisfies $\sum_{i=1}^nc_ii=0$ but $\sum_{i\in\{\alpha\}}
c_ii\neq0$ with $\{\alpha\}$ all proper subsets of consecutive
external lines. In addition, $\tilde{X}$ and $\eta_X$ are the
arbitrary reference spinor and Grassmann variable respectively. Thus
it follows from the corresponding recursion relation that the
N$^k$MHV super-amplitude can be expressed as the super-MHV
expansion, i.e.,
\begin{equation}
A_n=\sum_{\{\alpha_1\},\{\alpha_2\},\ldots,\{\alpha_k\}}
\frac{\delta^4(\sum_{i=1}^n\lambda_i\tilde{\lambda}_i)\delta^8(\sum_{i=1}^n\lambda_i\eta_i)}{\mathbf{cyc}(I_1)\mathbf
{cyc}(I_2)\ldots\mathbf{cyc}(I_{k+1})}\prod_{l=1}^k\frac{1}{P_{\alpha_l}^2}\prod_{A=1}^4[P_{\alpha_l}^2\eta_X^A+2\sum_{i\in\{\alpha_l\}}(i\tilde{X})\cdot
P_{\alpha_l}\eta_i^A],\label{CSW}
\end{equation}
where those internal line spinors implicit in $\mathbf{cyc}$ are
given by $\tilde{X}\cdot P_{\alpha_l}$.
\subsection{Proof of bonus relation for non-adjacent BCFW
deformations} For simplicity but without loss of generality, we
shall focus on the proof of bonus relation for the BCFW deformation
on the non-adjacent pair $[1,i]$. To achieve our goal, we firstly do
expand such a deformed amplitude on the basis of the BCFW recursion
relation for the pair $(1,2)$. Obviously, there are only two kinds
of diagrams contributing to the recursion relation, i.e., the
diagrams with $i$ staying with $1$ on the left or the diagrams with
$i$ staying with $2$ on the right. For the former case, the large
$z$ behavior is completely determined by the right hand side lower
point amplitude since the $z$ dependence comes only from this lower
point amplitude. What's more, such a $z$ dependence can be regarded
as the effect of the secondary BCFW deformation on the the
non-adjacent pair $[1,i]$ in the lower point on-shell amplitude. On
the other hand, for the latter case, the situation becomes a little
bit cumbersome, because the $z$ dependence comes from the three
parts, i.e., the propagator, the right and left hand side lower
point amplitudes. However, by the high school spinor analysis, one
can show the right and left hand side parts can be considered
effectively as the secondary super-BCFW deformation on the adjacent
pair $(1,P)$ and super-Risager deformation on the triple $\{i,P,2\}$
individually(Please refer to Appendix for explicit calculations). If
we choose $\tilde{1}$ and $\eta_1$ as the reference spinor and
Grassmann variable for the all-line supershifts (\ref{Risager}),
then by the corresponding super-MHV vertex expansion (\ref{CSW}),
the $z$ dependence of super-amplitude comes only from the
propagators under our triple super-Risager deformation. Therefore
the worst possible large $z$ behavior comes from the case where our
triple super-Risager deformation occurs on the same super-MHV
vertex, which gives us the order of $z^0$.

Now taking into account the fact that the large $z$ behavior for the
adjacent super-BCFW deformation and triple super-Risager deformation
go like $z^{-1}$ and $z^0$ respectively, we thus can prove that the
large $z$ behavior goes like $z^{-2}$ for the non-adjacent
super-BCFW deformation by induction. Note that the four point
scattering amplitude is just the MHV amplitude Eq.(\ref{MHV}), which
satisfies the $z^{-2}$ behavior for the non-adjacent super-BCFW
deformation, we thus complete our proof.
\section{Conclusion}
Along the line of S matrix program for massless spin-1 particles,
the consistency condition between the BCFW deformations on various
adjacent pairs gives us nothing but Lie algebra structure for the
coupling constant\cite{BC,HZ,ST}, which thus allows us to construct
the color stripped scattering amplitude by the BCFW deformation on
any adjacent pair.

On the other hand, the consistency condition and bonus relation
associated with the non-adjacent BCFW deformations give Kleiss-Kuijf
relation and BCJ relation on the scattering amplitude\cite{FHJ,JHL}.
It is now shown in this note that Kleiss-Kuijf relation and BCJ
relation follow the consistency condition and bonus relation for the
non-adjacent BCFW deformations respectively, which implies that such
a consistency condition and Kleiss-Kuijf relation are essentially
equivalent to each other, the same for the bonus relation and BCJ
relation.

Note that both the bonus relation and BCJ relation imply the
consistency condition and Kleiss-Kuijf relation. Therefore, we would
obtain a purely S matrix construction of these objects once we could
provide a proof of bonus relation or BCJ relation by the above
constructed scattering amplitude through any adjacent pair rather
than involve any field theory argument\cite{AK,BCJ}. As shown in
this note, we have virtually accomplished this task by reproducing
the bonus relation through a purely S matrix analysis in the context
of $\mathcal{N}=4$ SYM theory.

We conclude with one simple observation and two open problems.
Applying the same strategy given in Section $4$ to the adjacent BCFW
deformation on the pair $(1,n)$, it is obvious to obtain by
induction that the large $z$ behavior goes like $z^{-1}$ for such a
deformation, which thus provides an alternative way to argue for the
consistency condition among those adjacent BCFW
deformations\cite{HZ,ST}.

In addition, parallel to the bonus relation and BCJ relation in
gauge theory, gravity has the similar bonus relation and KLT
relation\cite{AK,KLT,TZ,BV,BDSV,BDHK}, so it is interesting to
investigate how the bonus relation and KLT relation are explicitly
related to each other in gravity theory. In particular, it is also
tempting to explore whether the bonus relation or KLT relation can
be obtained by a purely S matrix analysis.
\section*{Acknowledgements}
The author is indebted to Bo Feng and Song He for stimulating
discussions during this project. In addition, he would like to thank
Henriette Elvang and Michael Kiermaier for valuable communications
and suggestions. He is also grateful to Bianca Dittrich for the
hospitality during his stay at Albert Einstein Institute, where this
project was on progress. This project was partially supported by a
European Union grant FP7-REGPOT-2008-1-CreteHEP Cosmo-228644 and a
CNRS PICS grant \# 4172.
\appendix
\section*{APPENDIX}
Before we do the BCFW deformation on the non-adjacent pair $[1,i]$,
the deformation on the adjacent pair $(1,2)$ imposes the on-shell
condition on the internal line as
\begin{equation}
P^2-2z_0(2\tilde{1})\cdot P=0.\label{onshell1}
\end{equation}
Whence we can fix $z_0$ and the corresponding on-shell internal
momentum denoted by $(\lambda,\tilde{\lambda})$. Now for the later
convenience but without making difference in analysis of large $z$
behavior, we firstly do the BCFW deformation on the non-adjacent
pair $[1,i]$ as follows
\begin{eqnarray}
1(z)=1+z\langle\lambda|2\rangle i,\nonumber\\
\tilde{i}(z)=\tilde{i}-z\langle\lambda|2\rangle\tilde{1}.
\end{eqnarray}
Now by the BCFW deformation on the adjacent pair $(1,2)$, the
on-shell condition of the internal line gives us
\begin{equation}
P^2-2[(z'_02+z\langle\lambda|2\rangle i)\tilde{1}]\cdot
P=0.\label{onshell2}
\end{equation}
Setting $z'_0=z_0+z'$ and plugging Eq.(\ref{onshell1}) into
Eq.(\ref{onshell2}), we have
\begin{equation}
z'2+z\langle\lambda|2\rangle i=0.
\end{equation}
Then multiplying $\lambda$ yields
\begin{equation}
z'=z\langle i|\lambda\rangle.
\end{equation}
Furthermore, the momentum conservation for the right hand side
sub-amplitude also requires the shift of $\tilde{\lambda}$, i.e.,
\begin{equation}
\tilde{\lambda}(z)=\tilde{\lambda}-z\langle 2|i\rangle\tilde{1}.
\end{equation}
By the same token, the momentum conservation for the left hand side
sub-amplitude implies
\begin{equation}
1(z'_0)=1+z\langle\lambda|2\rangle i+z'_02=1+z_02+z\langle
2|i\rangle(-\lambda),
\end{equation}
which can also be obtained by Schouten identity indeed.

Note that the corresponding Grassmann variables are shifted in
accordance with the tilde spinors. Thus the $z$ dependence on the
sub-amplitudes can be regarded as the secondary super-BCFW
deformation on the adjacent pair $(1,P)$ on the left hand side
sub-amplitude and kind of super-Risager deformation on the tripe
$\{i,P,2\}$ on the right hand side one\cite{Risager}.


\begin{thebibliography}{10}
\bibitem{BCF}R. Britto, F. Cachazo, and B. Feng, Nucl. Phys. B715:
499(2005).
\bibitem{BCFW}R. Britto, F. Cachazo, B. Feng, and E. Witten, Phys.
Rev. Lett. 94: 181602(2005).
\bibitem{BC}P. Benincasa and F. Cachazo, arXiv:0705.4305[hep-th].
\bibitem{HZ}S. He and H. Zhang, arXiv:0811.3210[hep-th].
\bibitem{ST}P. C. Schuster and N. Toro, JHEP 0906: 079(2009).
\bibitem{FHJ}B. Feng, R. Huang, and Y. Jia, arXiv:1004.3417[hep-th].
\bibitem{AK}N. Arkani-Hamed and J. Kaplan, JHEP 0804: 076(2008).
\bibitem{KK}R. Kleiss and H. Kuijf, Nucl. Phys. B312: 616(1989).
\bibitem{BCJ}Z. Bern, J. J. M. Carrasco, and H. Johansson, Phys. Rev. D78: 085011(2008).
\bibitem{DHKS}J. M. Drummond, J. Henn, G. P. Korchemsky, and E.
Sokatchev, Nucl. Phys. B828: 317(2010).
\bibitem{ACK}N. Arkani-Hamed, F. Cachazo, and J. Kaplan,
arXiv:0808.1446[hep-th].
\bibitem{Nair}V. P. Nair, Phys. Lett. B214: 215(1988).
\bibitem{EFK1}H. Elvang, D. Freedman, and M. Kiermaier, JHEP 0904:
009(2009).
\bibitem{EFK2}H. Elvang, D. Freedman, and M. Kiermaier, JHEP 0906:
068(2009).
\bibitem{KN}M. Kiermaier and S. G. Naculich, JHEP 0905: 072(2009).
\bibitem{JHL}Y. Jia, R. Huang, and C. Liu, arXiv:1005.1821.
\bibitem{KLT}H. Kawai, D. C. Lewellen, and S. H. H. Tye, Nucl. Phys. B269: 1(1986).
\bibitem{TZ}S. H. H. Tye and Y. Zhang, arXiv:1003.1732[hep-th].
\bibitem{BV}N. E. J. Bjerrum-Bohr and P. Vanhove,
arXiv:1003.2396[hep-th].
\bibitem{BDSV}N. E. J. Bjerrum-Bohr, P. H. Damgaard, T. Sondergaard, and P. Vanhove,
arXiv:1003.2403[hep-th].
\bibitem{BDHK}Z. Bern, T. Dennen, Y. Huang, and M. Kiermaier, arXiv:1004.0693[hep-th].
\bibitem{Risager}K. Risager, JHEP 0512: 003(2005).
\end{thebibliography}
\end{document}